\documentclass[a4paper,11pt]{article}
\pdfoutput=1 

\usepackage{jinstpub} 

\usepackage[version=3]{mhchem}
\usepackage{siunitx}
\usepackage{natbib}
\usepackage{amsmath} 

\usepackage{verbatim}
\usepackage{multirow}

\usepackage[english]{babel}
\usepackage{hyperref}
\addto\extrasenglish{
    
}

\usepackage{setspace}
\usepackage{graphicx}
\usepackage{amsmath}
\usepackage{siunitx}

\usepackage{pbox}
\usepackage{xcolor}

\newcolumntype{"}{@{\hskip\tabcolsep\vrule width 1pt\hskip\tabcolsep}}
\DeclareSIUnit[number-unit-product = \;]\year{yr}
\DeclareSIUnit\parsec{pc}
\DeclareSIUnit\torr{Torr}
\DeclareSIUnit\centimeter{\centi \meter}
\DeclareSIUnit\sq{\ensuremath{\Box}}
\usepackage{textcomp}
\usepackage{caption}
\usepackage{subcaption}
\usepackage{url}
\usepackage{ctable}

\title{Charge Amplification in Sub-atmospheric CF$_4$:He Mixtures for Directional Dark Matter Searches}

\author[a,1]{A.G. McLean,\note{Corresponding author.}}
\author[a]{N.J.C. Spooner,}
\author[b]{T. Crane,}
\author[a]{C. Eldridge,}
\author[a]{A.C. Ezeribe,}
\author[a,c]{R.R. Marcelo Gregorio,}
\author[a]{and A. Scarff}

\affiliation[a]{Department of Physics and Astronomy, University of Sheffield, South Yorkshire, S3 7RH, United Kingdom}
\affiliation[b]{AWE plc, Aldermaston, Reading, Berkshire, RG7 4PR, United Kingdom}
\affiliation[c]{School of Physical and Chemical Sciences, Queen Mary University of London, E1 4NS, United Kingdom}

\emailAdd{ali.mclean@sheffield.ac.uk}

\abstract{Low pressure gaseous Time Projection Chambers (TPCs) are a viable technology for directional Dark Matter (DM) searches and have the potential for exploring the parameter space below the neutrino fog \cite{Vahsen2020, OHare2021}. Gases like CF$_4$ are advantageous because they contain flourine which is predicted to have heightened elastic scattering rates with a possible Weakly Interacting Massive Particle (WIMP) DM candidate \cite{Ellis1991, Bednyakov, Divari}. The low pressure of CF$_4$ must be maintained, ideally lower than 100 Torr, in order to elongate potential Nuclear Recoil (NR) tracks which allows for improved directional sensitivity and NR/Electron Recoil (ER) discrimination \cite{Phan2016}. Recent evidence suggests that He can be added to heavier gases, like CF$_4$, without significantly affecting the length of $^{12}$C and $^{19}$F recoils due to its lower mass. Such addition of He has the advantage of improving sensitivity to lower mass WIMPs \cite{Vahsen2020}. Simulations can not reliably predict operational stability in these low pressure gas mixtures and thus must be demonstrated experimentally. In this paper we investigate how the addition of He to low pressure CF$_4$ affects the gas gain and energy resolution achieved with a single Thick Gaseous Electron Multiplier (ThGEM).}


\begin{document}
\maketitle
\flushbottom
\raggedbottom

\section{Introduction}
\label{sec:intro}

Dark Matter (DM) comprises 85\% of the mass in the observable Universe. The existence of Weakly Interacting Massive Particles (WIMPs) is one of the possibilities that could transpire to account for DM, yet the direct measurement of such a particle has not been confirmed. In the field of DM detection, many attempts have been made to measure the rare low energy events expected when WIMPs elastically scatter off nuclei; despite significant improvements in the sensitivity of WIMP detectors in recent years, the positive identification of recoils instigated by WIMPS has not been established \cite{Billard2022}. Results from the leading two-phase xenon Time Projection Chamber (TPC) experiments, LZ and XENON, indicate that there is no excess in the number of events close to the neutrino fog \cite{LZ,XENON}. The future generations of these detectors will struggle to discriminate between Nuclear Recoils (NRs) induced by neutrinos, predominately coming from the Sun, and NRs induced by WIMPs \cite{OHare2021, Billard2014}.

Discrimination between NRs induced by WIMPs and Solar neutrinos is theoretically possible by utilising low pressure gaseous TPCs; the direction of a NR can be determined by reconstructing the track of ionisation left behind in its wake. Such a measurement would provide a Galactic signature which can be used as a means for discrimination. Solar neutrinos could be distinguished from WIMP signals which, due to the motion of the Solar System around the Galaxy, are expected to originate from the direction of the Cygnus constellation. This Galactic signature is further modulated by the motion of the Earth around the Sun \cite{Spergel1988,Copi1999} and by the rotation of the Earth on its own axis which causes a directional shift over the course of a sidereal day \cite{Morgan2003}. 
Unlike an annual modulation experiment, a DM detector with directional capability would be able to measure a Galactic signal with confidence as it could not easily be replicated by a terrestrial source \cite{Klinger2015,Kudryavtsev2010,Davis2014}.

The DRIFT detector currently holds the best published sensitivity for directional DM searches by utilising a CS$_2$:CF$_4$:O$_2$ mixture at the low pressure of 41 Torr (30:10:1 Torr) \cite{Bat2016}. These low pressures are required in order to facilitate longer NR tracks on the millimeter scale. Since DRIFT's latest published search results, gases containing fluorine like CF$_4$ have become more popular as the $^{19}$F content is suspected to improve the spin-dependent elastic scattering rates with a possible WIMP candidate \cite{Ellis1991,Bednyakov, Divari}. 
Additionally, CF$_4$ is proven to be readily capable of producing sufficient gas gains with a single Thick Gaseous Electron Multiplier (ThGEM) at low pressures \cite{Burns2017,Callum_thesis}.

One challenge of using low pressure CF$_4$ gas is the resulting low target mass. In order to probe meaningful cross sections such a low pressure detector would need to be $\mathcal{O}$(10 - 100) cubic meters \cite{Ahlen2010, Deaconu2017}. This issue can not be addressed by simply increasing the pressure without also sacrificing the directional sensitivity of the detector. Several studies have shown that the recoil length of $^{12}$C and $^{19}$F are on the order of millimeters in low pressure CF$_4$ \cite{Deaconu2017, Deaconu_thesis, Phan2016, Miuchi2007, Ahlen2011}; including the successful observation of the head tail effect \cite{Dujmic2008, Scarff}, which is important for determining the direction along the principle axis of a recoil.  It has also been found that the separation of NR and Electron Recoil (ER) populations at lower energy thresholds could be improved by reducing the pressure of CF$_4$ below 100 Torr \cite{Phan2016}.

To address this low pressure limitation one possibility could be to add He to gases like CF$_4$. The addition of He would increase the target mass and improve sensitivity to lower WIMP masses, without having a detrimental effect on the recoil track length of $^{12}$C and $^{19}$F nuclei due to its much lower density \cite{Vahsen2020}. A number of researchers have already begun testing these He mixtures \cite{Fraga2003, Margato, Vahsen2014, Coimbra2016, Baracchini2018, Baracchini2020, Amaro2023}. Atmospheric mixtures of CF$_4$:He in molar ratios of 40:60 and 20:80 have already been successfully demonstrated \cite{Baracchini2020, Vahsen2014}. Sub-atmospheric CF$_4$:He mixtures with a molar ratio of 20:80 have also been tested before with a double GEM configuration but the gas mixture could not be operated at pressures lower than 400 Torr \cite{Vahsen2014}. Additionally, these measurements were done with a premixed gas bottle, so the partial pressure of CF$_4$ could not be managed independently of total pressure; as discussed, the partial pressure of CF$_4$ should take priority to ensure directionality of both $^{12}$C and $^{19}$F recoils.

In this paper, efforts to evaluate the charge amplification performance of CF$_4$:He mixtures at sub-atmospheric pressures, while also prioritising constant low partial pressure of CF$_4$, are presented. To begin, a description of the ThGEM TPC used and methodology is discussed along with the gas filling procedure. Following this, gas gain curves and energy resolution results are presented for both pure CF$_4$ and CF$_4$:He mixtures at pressures which could produce a significant gas gain. Finally, the results are concluded and possible avenues for future work are discussed.

\section{Single ThGEM TPC Setup and Methodology}
\label{sec:design}

\begin{figure} [h]
     \centering
     \captionsetup[subfigure]{justification=centering}
     \begin{subfigure}[h]{0.45\textwidth}
         \centering
         \includegraphics[height = 7cm]{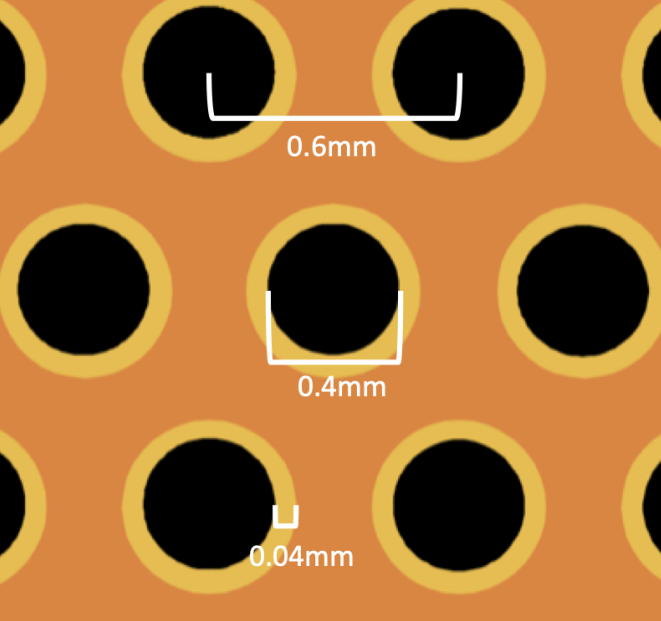}
         \caption{}
         \label{fig:ThGEM_top_down}
     \end{subfigure}
     \hfill
     \begin{subfigure}[h]{0.45\textwidth}
         \centering
         \includegraphics[angle=270,origin=c,trim={26cm 0cm 10cm 0cm},clip, height=7cm]{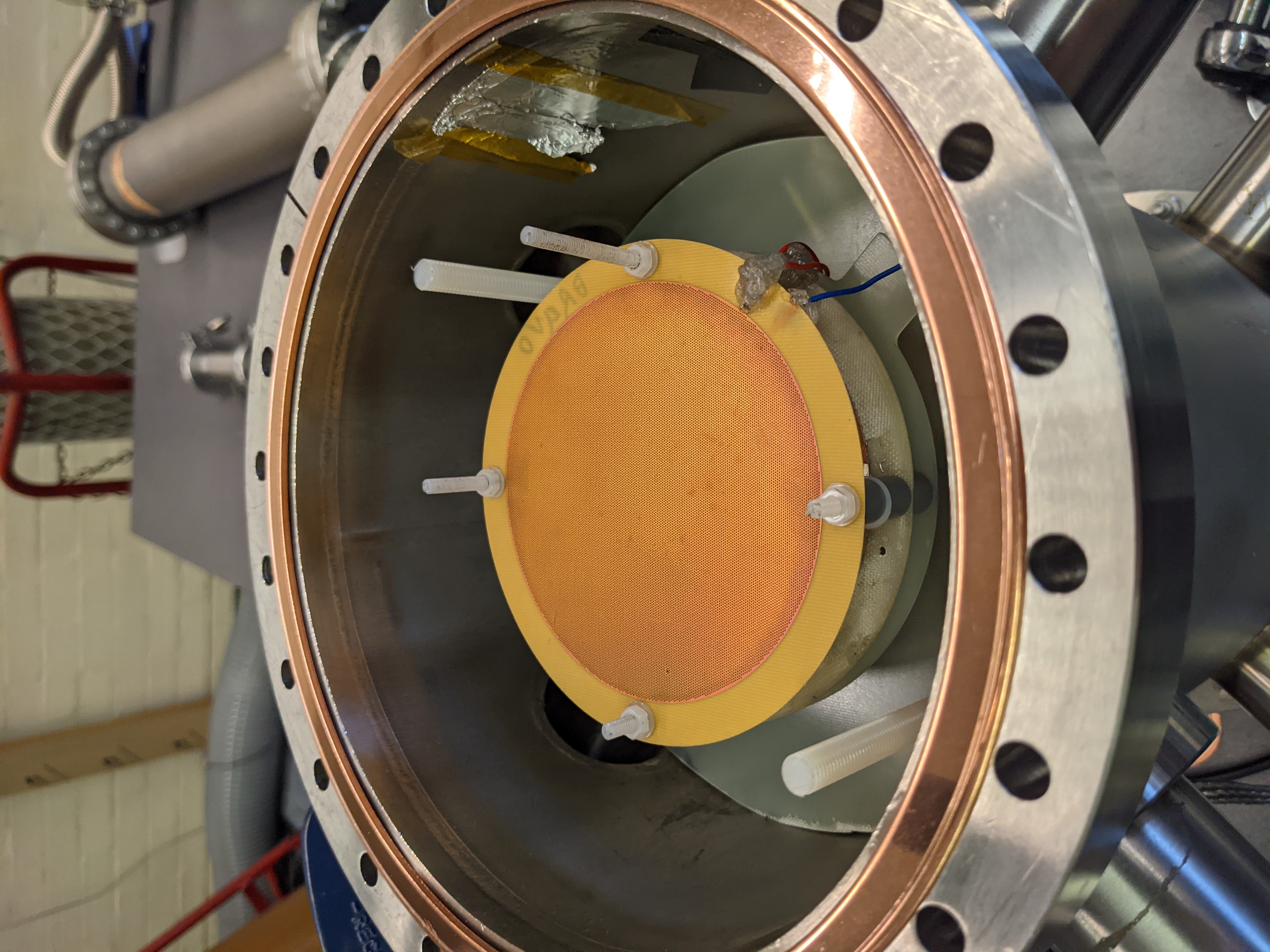}
         \caption{}
         \label{fig:experiment_image}
     \end{subfigure}
    \caption{Magnified top down diagram depicting the dimensions of the hole structures of the ThGEM \cite{Callum_thesis} (left). Image of small scale TPC assembly inside the open vacuum vessel (right).}
    \label{fig:three graphs}
\end{figure}

The ThGEM used in this work, denoted as ThGEM-B previously in \cite{Callum_thesis}, was produced at CERN. The device has a thickness of 0.4 mm and, as can be seen in \autoref{fig:ThGEM_top_down}, the holes have a diameter of 0.4 mm and a pitch of 0.6 mm. The ThGEM top and bottom Cu layers around the holes have been etched in such a way to create a rim structure. This rim is important primarily because it reduces the probability of electrical discharge \cite{Chechik2006, Bressler2014}. Moreover, charge which collects on the surface of the rim helps to shape the field close to the holes \cite{Pitt2018}. The rim of the holes of this particular ThGEM have a width of 0.04 mm.

This ThGEM was then mounted with a cathode to a fibre glass base and was placed inside a small vacuum vessel. An image of this mounting setup can be seen in \autoref{fig:experiment_image}. The cathode was offset by 1.5cm above the base and the ThGEM was then offset by 2 cm above the cathode. The cathode was biased with a negative High Voltage (HV), the bottom plane of the ThGEM was grounded, and the top plane was biased with a positive HV via the biasing line of a CREMAT CR-150 evaluation board. The potential difference between the ThGEM top and bottom planes is defined as \(\Delta V_{\text{ThGEM}}\). Both HV channels were provided by an NHQ 202M iseg HV supply. The Evaluation board utilised a CREMAT CR-111 charge sensitive preamplifier chip. The exposure of the TPC assembly to an $^{55}$Fe X-ray source was controlled by securing a magnet to the source and then using a secondary magnet externally to hold the source at the desired height. This assembly was then enclosed in the vessel and evacuated for a minimum of 48 hours using a vacuum scroll pump and achieved a vacuum < 10$^{-2}$ Torr. The vessel had a leak rate < 0.1 Torr per day.

The output of the preamplifier was connected externally to a CREMAT CR-200-4$\mu$s shaper module on a CREMAT CR-160 shaper evaluation board. The output of the shaper was connected to an Ortec 926 ADCAM MCB which interfaced with the software package Maestro \cite{Maestro}. This allowed the signals to be binned according to their amplitude. The charge sensitive electronics were calibrated by injecting test pulses from an Ortec 480 Pulser into a 1 pF capacitor on the CREMAT CR-150 evaluation board. For all presented results, the ADC bin number was calibrated against gas gain via the W-value of pure CF$_4$, 34 eV \cite{Reinking}, and energy of the $^{55}$Fe X-ray, 5.89 keV, by varying the amplitude of the test pulses. This is a valid approximation for the CF$_4$:He mixtures as previous weighted calculations have shown that the W-value of CF$_4$ dominates that of He, even despite being a minority component \cite{Vahsen2014}. The variation in the number of initial electron-ion pairs across all mixtures is expected to be $\sim$ 5 pairs.

\begin{figure} [h!]
    \captionsetup[subfigure]{justification=centering, labelformat=empty}
    \centering
    \begin{subfigure} {.5\textwidth}
        \centering
      \includegraphics[width=\textwidth]{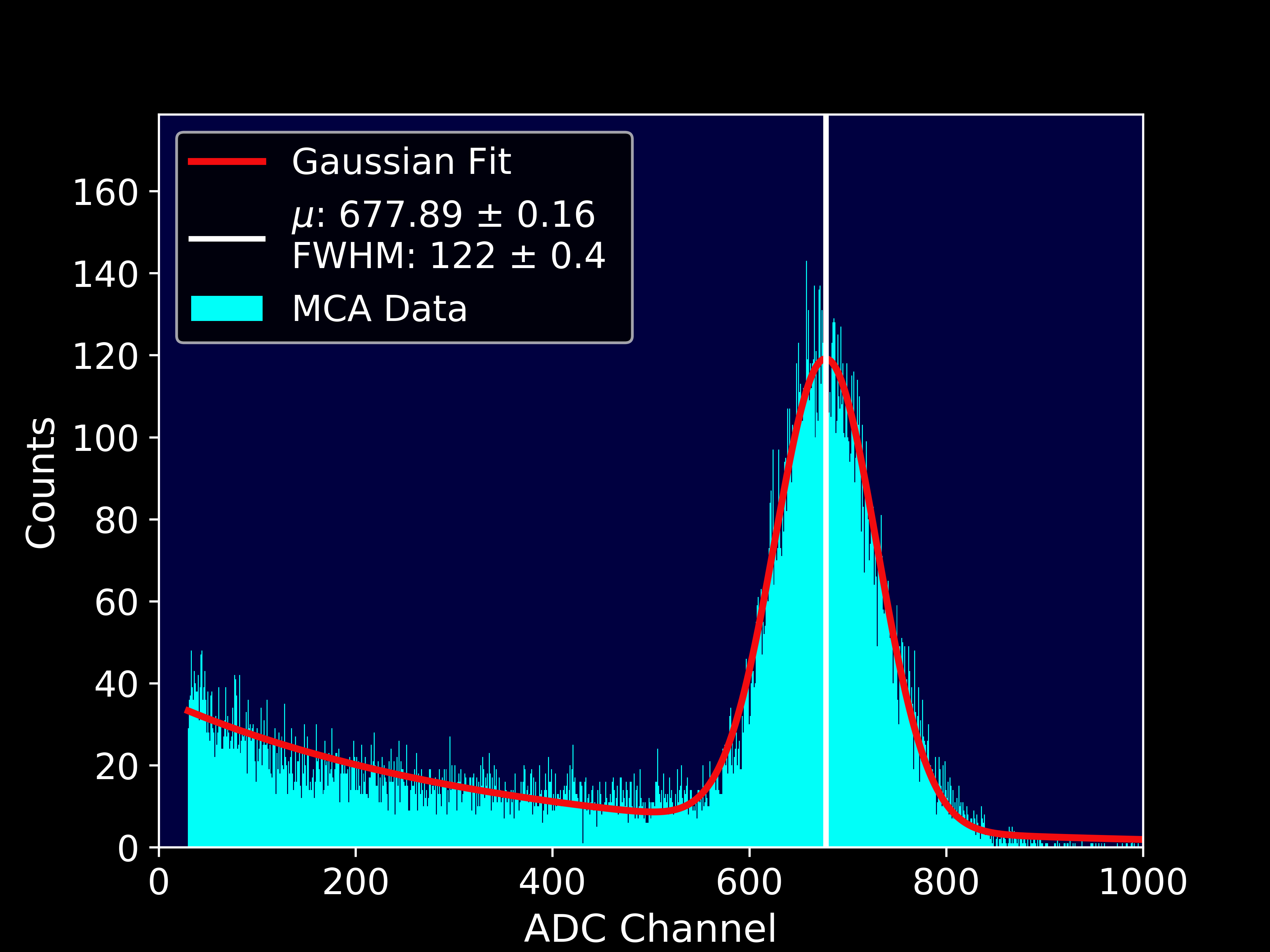}
      \caption{}
      \label{fig:ex_max_spec}
    \end{subfigure}%
    \begin{subfigure} {.5\textwidth}
      \centering
      \includegraphics[width=\textwidth]{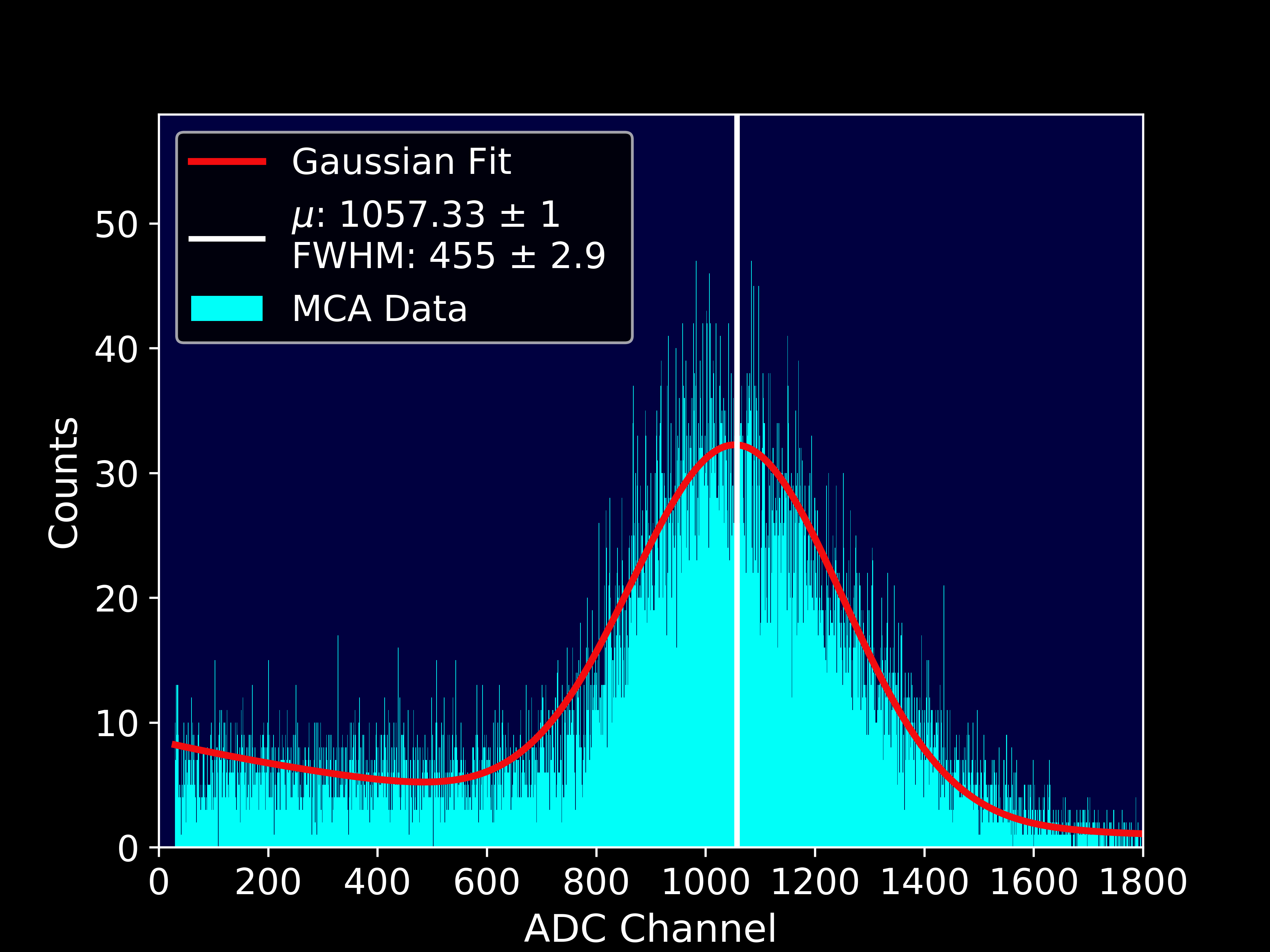}
      \caption{}
      \label{fig:ex_int_spec}
    \end{subfigure}%
    \vspace{-1.4\baselineskip}
    \caption{Example $^{55}$Fe spectrum in 40 torr of pure CF$_4$ with \( \Delta V_{\text{ThGEM}}\) = 640 V (left). Example $^{55}$Fe spectrum in a CF$_4$:He mixture with partial pressures 40:720 Torr with \( \Delta V_{\text{ThGEM}}\) = 740 V (right).}
    \label{fig:spec}
\end{figure}

Following an exposure to the $^{55}$Fe source, energy spectra were obtained, examples of which can be seen in \autoref{fig:spec}. The gain was then extracted by fitting a gaussian and an exponentially falling component to the spectrum. The mean of the gaussian was used to determine the gas gain via the calibration of the electronics. The energy resolution was then extracted by calculating the FWHM of the gaussian and then dividing by the mean. The spectrum on the left in \autoref{fig:spec} shows an example with a relatively small energy resolution, conversely the spectrum on the right shows an example with a relatively large energy resolution. 

\begin{figure}[b]
    \centering
    \includegraphics[width=\textwidth,trim={0.2cm 0cm 0cm 0.2cm},clip]{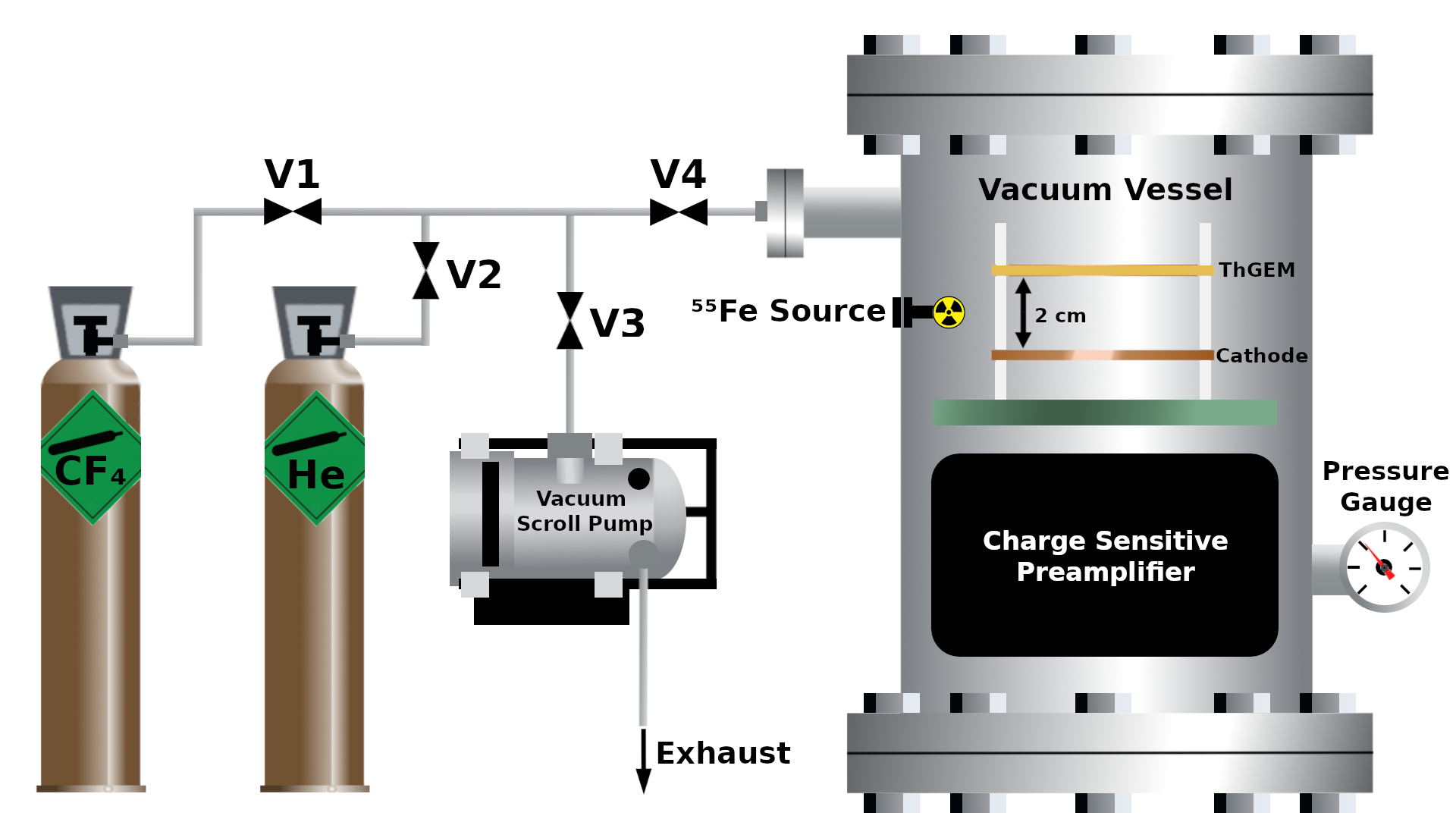}
    \caption{Diagram of the experimental setup and gas system used to evacuate and fill the vacuum vessel with CF$_4$ and CF$_4$:He mixtures to the desired partial pressures.}
    \label{fig:Experimental_Setup}
\end{figure}

When the vessel was ready to be filled with the target gas mixture, a filling procedure was followed in a attempt to mix the gas in a repeatable way. \autoref{fig:Experimental_Setup} shows a diagram which aids the explanation of the filling procedure. During the evacuation phase: the gas bottles were closed, the pump was turned on and valves V1 - V4 were all opened. The filling procedure begins by closing all the valves and turning off the pump. Both gas bottles were then briefly opened and closed to reduce the chance of gas contaminants leaking into the gas lines due to a large pressure differential. Starting with CF$_4$, the gas bottle and V1 were opened. V4 was then used as a throttle to fill the vacuum vessel to the desired pressure in a slow and controlled manner. Once the desired pressure of CF$_4$ (\(p_{CF_4}\)) was reached on the pressure gauge, V4 was closed along with V1 and the gas bottle. If only pure CF$_4$ was required for a measurement then this concluded the filling procedure. However when a mixture was required, the pump was then turned back on and V3 was opened to evacuate the gas line for 5 minutes. Afterwards, V3 was closed and the pump was turned off again. Then the He gas bottle and V2 were opened and V4 was again used as a throttle to fill the vessel to the desired partial pressure. Once at the correct pressure, both valves were closed along with the gas bottle. Finally, after the measurements were complete, the gas system was returned to the evacuation phase.

\section{Low Pressure Pure CF$_4$}
\label{sec:PureCF4}

\begin{figure}[b]
    \centering
    \includegraphics[width=0.9\textwidth]{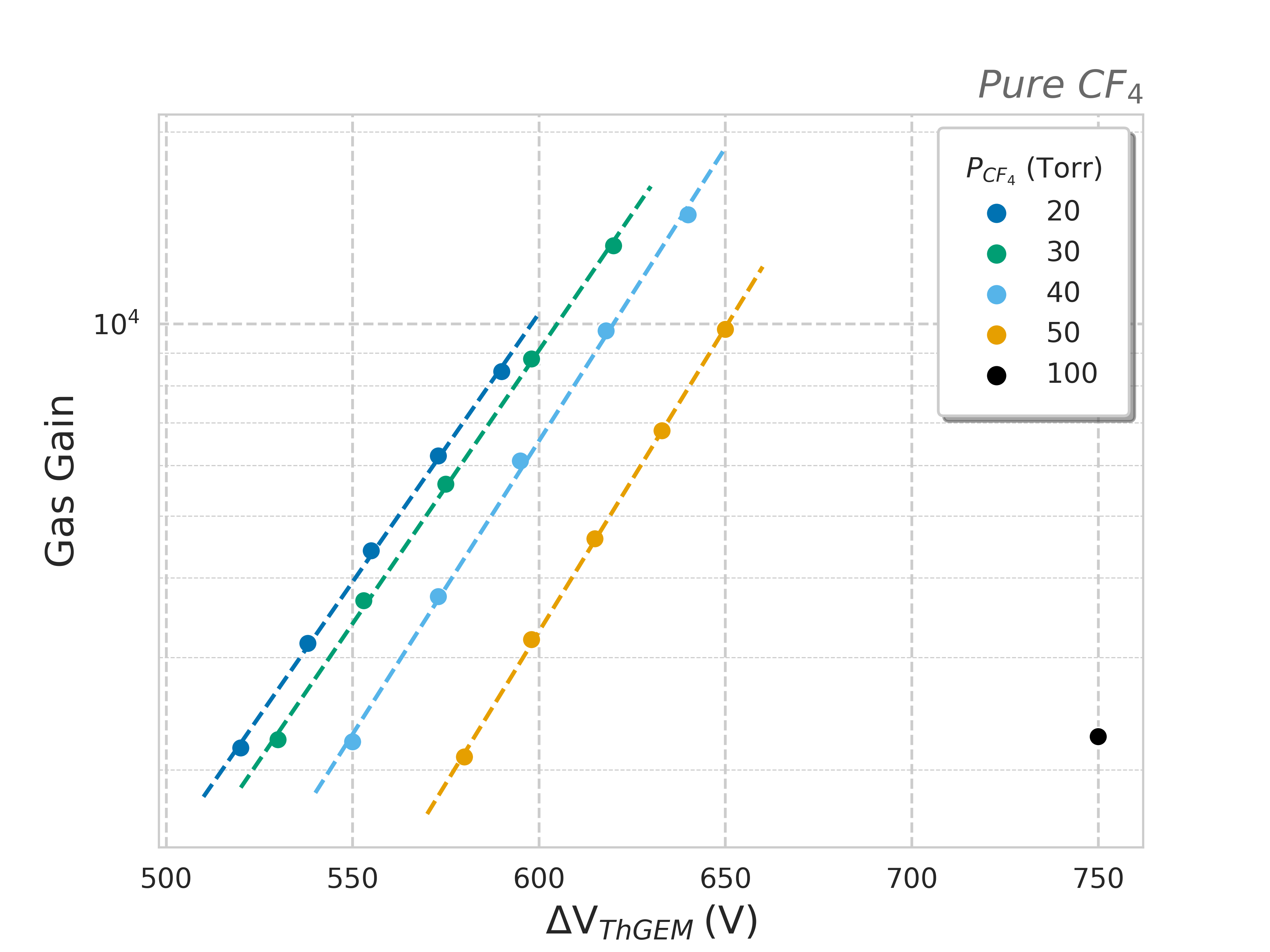}
    \caption{Gas gains measured in pure CF$_4$ at a range of low pressures presented on a log scale.}
    \label{fig:GG_pure_CF4}
\end{figure}

As mentioned, the pressure of CF$_4$ should be low and ideally lower than 100 Torr in order to improve directional sensitivity and NR/ER discrimination. For this reason, a range of low pressures of pure CF$_4$ were tested; 5, 10, 20, 30, 40, 50, and 100 Torr. The filling procedure, as described in \autoref{sec:design}, was followed for each pressure. 

Once the vessel was filled, the cathode voltage was set to -300 V; any possible systematic effects regarding the reduced drift field strength are assumed to be negligible. Then the ThGEM top voltage was increased until a clear $^{55}$Fe photo-peak could be observed in the energy spectrum. Following this, the voltage was increased gradually in increments of 10 V until sparking occurred between the ThGEM top and bottom; indicated by the tripping of the current limiter on the HV supply connected to the ThGEM top. This allowed a range of operational voltages to be determined for a given pressure. Once this range was determined, the upper voltage limit was set by reducing the voltage by 10 V below the sparking limit. Intermediate voltage settings were determined by dividing the operational voltage range into five sample voltages. The result of these gain measurements can be seen in \autoref{fig:GG_pure_CF4}.

As can be seen in \autoref{fig:GG_pure_CF4}, gain curves were obtained for 20 - 50 Torr of pure CF$_4$ and exhibit an exponential increase in gas gain with increasing voltage. Accordingly, exponential curves were fitted for each pressure and are displayed as dashed lines. Only a single data point was captured at 100 Torr due to a very narrow range of operating voltages before sparking occurred. Gain measurements could not be established in 5 and 10 Torr of pure CF$_4$ because sparking was observed before a clear photo-peak could be seen on the energy spectrum. For the observed gain curves, it can be seen that as the pressure of CF$_4$ increases, larger  \( \Delta V_{\text{ThGEM}}\) voltages were required to achieve the same gas gain. This causes the gain curves to shift to the right with increasing pressure. The 40 and 50 Torr curves in \autoref{fig:GG_pure_CF4} are consistent with previous measurements conducted with ThGEMs with identical dimensions \cite{Callum_thesis, Burns2017}. However, the maximum gas gains achieved in \autoref{fig:GG_pure_CF4} are smaller than those previous results and could be a result of sparking damage discussed in \cite{Callum_thesis}.

\begin{figure}[b]
    \centering
    \includegraphics[width=0.9\textwidth]{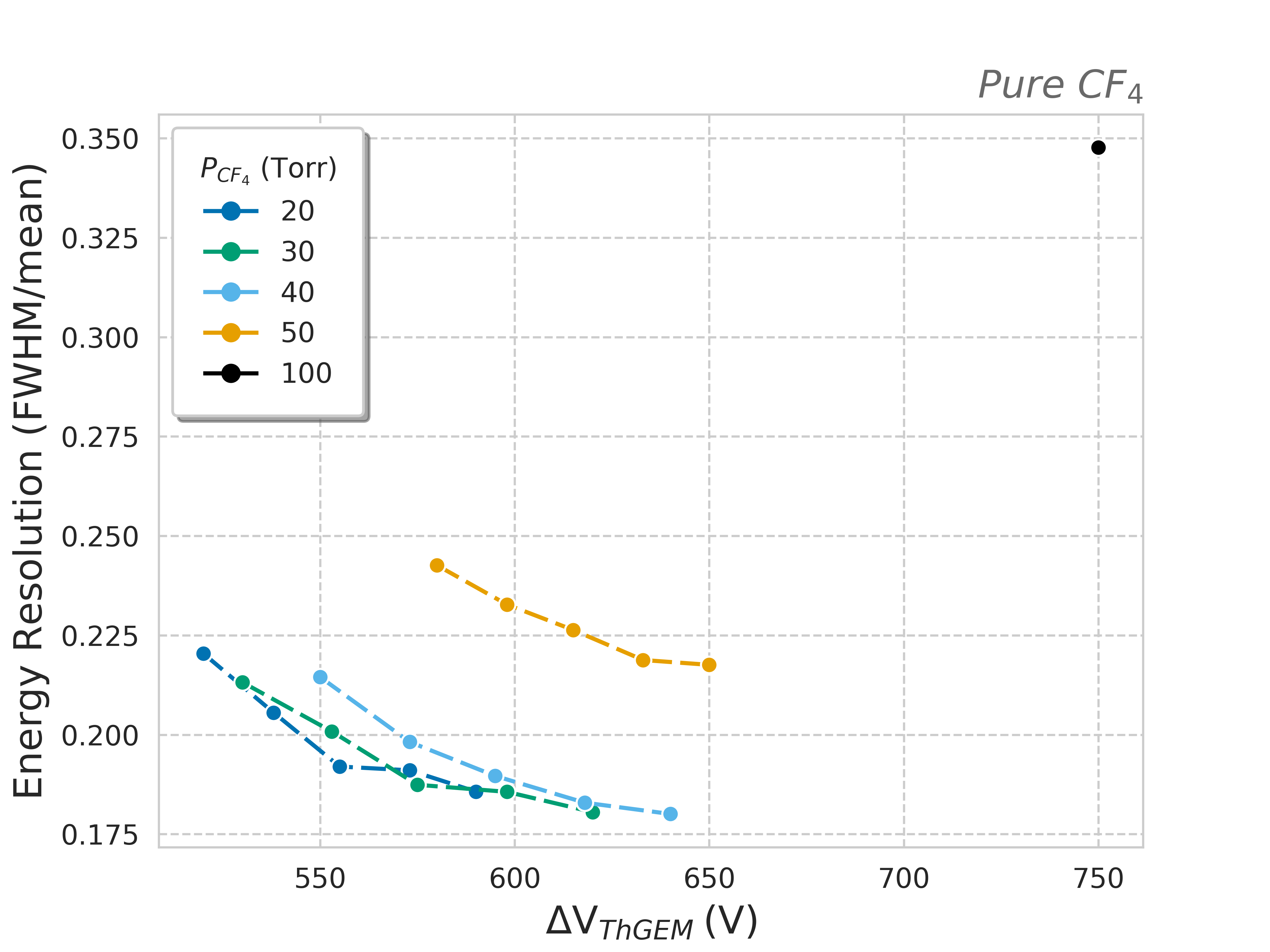}
    \caption{Energy resolution measured in pure CF$_4$ at a range of low pressures.}
    \label{fig:ER_pure_CF4}
\end{figure}

The energy resolution was also calculated for each pressure and can be seen presented as a function of  \( \Delta V_{\text{ThGEM}}\) in \autoref{fig:ER_pure_CF4}. The figure shows that the energy resolution decreases with increasing voltage. For a given potential difference, the energy resolution generally increases with increasing pressure. The energy resolution curves of 20, 30 and 40 Torr are very similar; however at 50 Torr the energy resolution increases significantly. The best energy resolution in 50 Torr is comparable to the worst energy resolutions measured in 20, 30 and 40 Torr. Furthermore, the energy resolution measured at 100 Torr is significantly worse than all the measurements at lower pressure. This suggests that, in addition to being beneficial for directionality, lowering the pressure of CF$_4$ below 100 Torr improves the energy resolution. 

The best energy resolution, 0.1801 $\pm$ 0.0006, was achieved during the 40 Torr run and occurred at a  \( \Delta V_{\text{ThGEM}}\) of 640 V. Conversely, the worst energy resolution was measured to be 0.3477 $\pm$ 0.0007 at 100 Torr with a \( \Delta V_{\text{ThGEM}}\) of 750 V. In the following section, we will discuss how the addition of He affects the gain and energy resolution.

\section{Sub-atmospheric CF$_4$:He Mixtures}

\begin{figure}[t!]
     \centering
     \includegraphics[width=\textwidth,trim={1.8cm 2cm 3.1cm 4.2cm},clip]{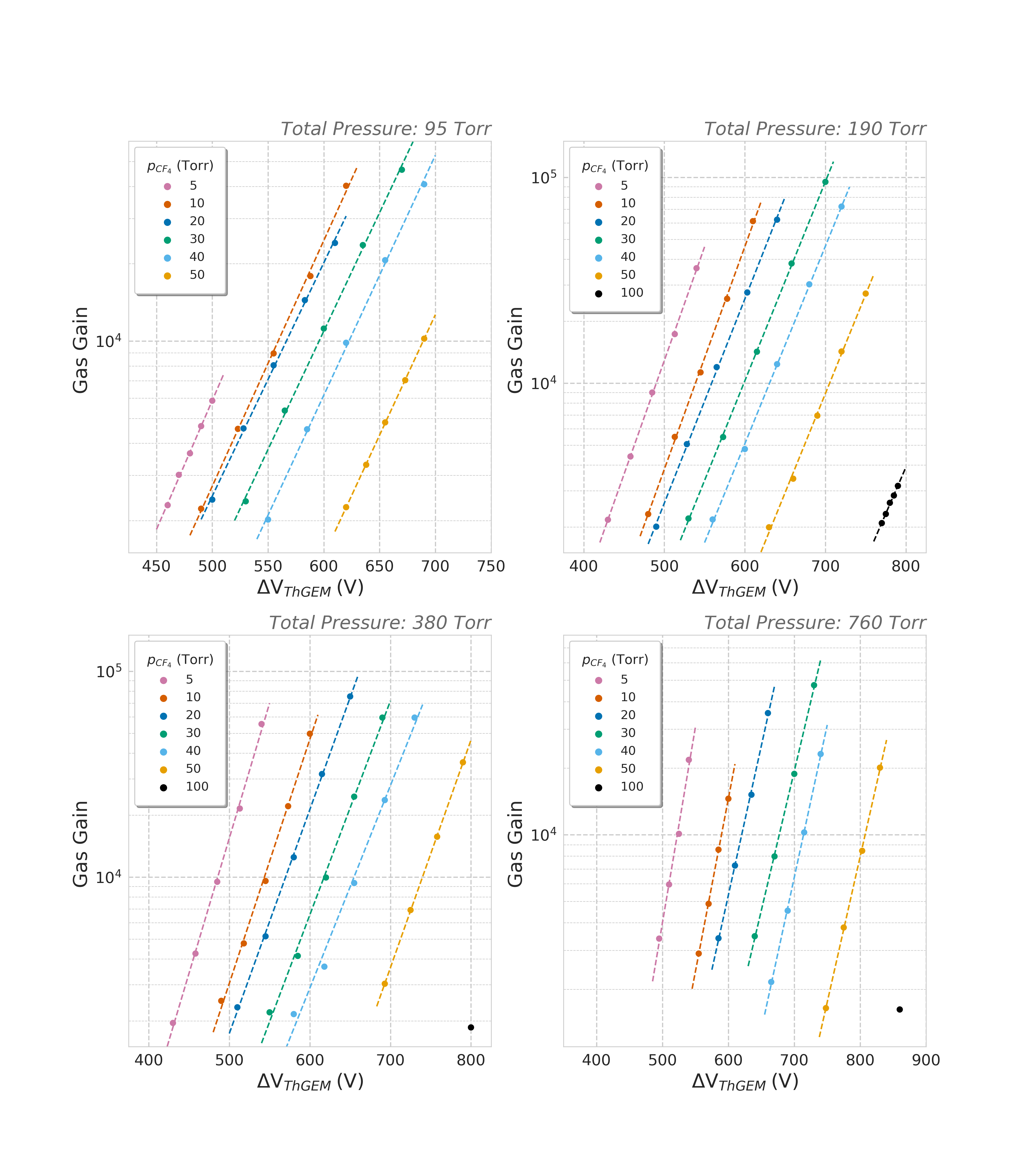}
     \caption{Gas gains measured in CF$_4$:He mixtures grouped together by total pressure presented on log scales.}
     \label{fig:He_gas_gains}
 \end{figure}

After the measurements in pure CF$_4$, He was added to the vessel using the filling procedure described in \autoref{sec:design}. For each CF$_4$ pressure, He was added gradually and gas gain measurements were taken at total pressures of 95, 190, 380, and 760 Torr. These pressures were chosen to give a broad range of sub-atmospheric pressures equivalent to 1/8, 1/4, 1/2, and 1 atm respectively. The same procedure for determining the operational voltage range, as described in \autoref{sec:PureCF4}, was used. The subsequent gas gain measurements can be seen in \autoref{fig:He_gas_gains}.

The gain curves in \autoref{fig:He_gas_gains} are grouped together with mixtures of equal total pressure; 95, 190, 380, 760 Torr from left-to-right, top-to-bottom. The colour represents the partial pressure of CF$_4$. Gain curves exhibiting exponential behaviour were obtained for all gas mixtures except in the case of 100 Torr. The gain curves follow a similar trend to that observed in the pure CF$_4$ measurements, i.e. as the partial pressure of CF$_4$ increases the \( \Delta V_{\text{ThGEM}}\) required to achieve the same gas gain increases. This trend is consistent for each given total pressure.

As can be seen, the addition of helium to CF$_4$ appears to stabilise the gas by increasing the maximum attainable gas gain before breakdown occurs in the gas. Previously in \autoref{sec:PureCF4}, sparking was observed in both 5 and 10 Torr of pure CF$_4$ before a photo-peak could be resolved above the noise. However with the addition of He, all mixtures with a CF$_4$ partial pressure of 5 and 10 Torr were able to yield gain curves. Furthermore, the operational voltage range with 100 Torr of CF$_4$ appears to increase with the addition of He as a range of voltages can be seen at a total pressure of 190 Torr in \autoref{fig:He_gas_gains}. As the partial pressure of He increases further, the operational voltage range is limited again and only one stable voltage could be measured at 380 and 760 Torr. This dramatic change in gas gain and stability for a given partial pressure of CF$_4$ is unsurprising as the mixing ratio is changing with increasing partial pressure of helium.


This improvement in gas gain is seen across all partial pressures of CF$_4$. If the maximum gas gains achieved in \autoref{fig:GG_pure_CF4} are compared with those in \autoref{fig:He_gas_gains} at 95 and 190 Torr, it can be seen that the addition of He has significantly improved the maximum stable gas gain. This improvement is almost an order of magnitude in some cases. At higher total pressures, 380 and 760 Torr, it can be seen that the maximum stable gas gains begin to recede. It is interesting to note however, that the gas gains at atmospheric pressure are still larger than their low pressure pure CF$_4$ counterpart. This could be due in part to He$^*$/CF$_4$ penning ionisation because the ionisation potential of CF$_4$ is smaller than the excitation potential of He. Further work is required in order to understand the penning nature of these mixtures.

\begin{figure}[t!]
    \centering
    \includegraphics[width=\textwidth,trim={1.8cm 2cm 3.1cm 4.2cm},clip]{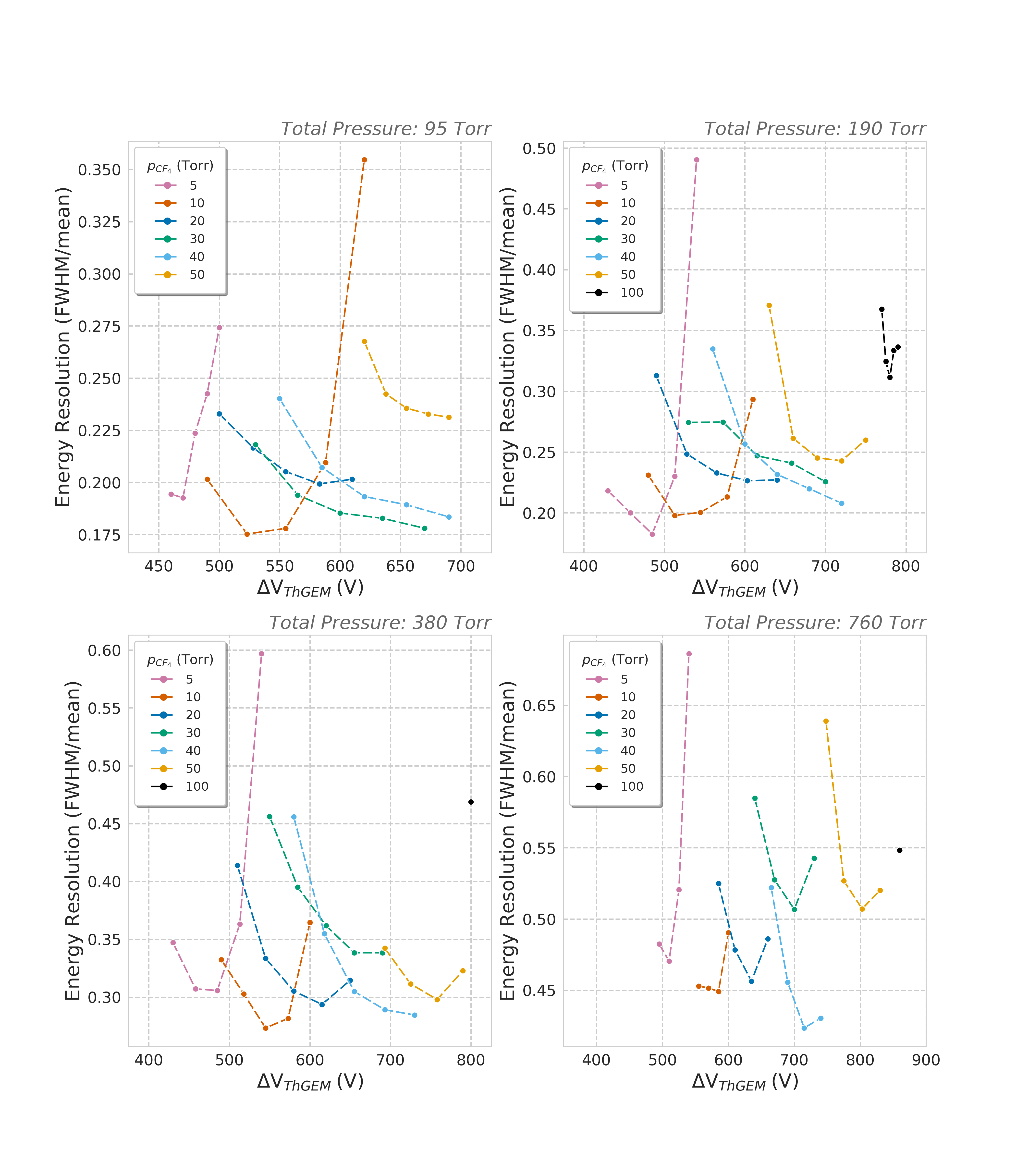}
    \caption{Energy resolution measured in CF$_4$:He mixtures grouped together by total pressure.}
    \label{fig:He_ER}
\end{figure}

The energy resolution of these CF$_4$:He mixtures was also evaluated. These results can be seen in \autoref{fig:He_ER}. If we consider first the mixtures with a CF$_4$ partial pressure of 20 - 50 Torr, we see that the energy resolution generally decreases with increasing \( \Delta V_{\text{ThGEM}}\) at a total pressure of 95 and 190 Torr. As the total pressure increases to 380 and 760 Torr, the energy resolution in these curves initially decreases with increasing \( \Delta V_{\text{ThGEM}}\) before exhibiting a slight increase at the highest operating voltages. This produces a minimum on many of the curves coinciding with a relatively high gas gain which, if so desired, could be used for \( \Delta V_{\text{ThGEM}}\) optimisation. Energy resolution curves with minima have been observed in CF$_4$ mixtures during ThGEM measurements before. This has previously been attributed to the shape and strength of the field lines extending past the top and bottom planes of the ThGEM. This is suspected to cause degradation in the energy resolution via photon feedback and variation in electron pathways through the ThGEM \cite{Coimbra}.

Local minima can also be seen in the mixtures containing a CF$_4$ partial pressure of 5 and 10 Torr; however, these occur at the lower end of the operating voltage range and would therefore not be ideal for optimising both energy resolution and gain simultaneously. A similar optimisation could be made for the mixture at a total pressure of 190 Torr with a CF$_4$ partial pressure of 100 Torr. Although this would not be very valuable as the gas gains in this case were relatively small compared to the lower CF$_4$ partial pressures, as seen in \autoref{fig:He_gas_gains}.

It can also be seen that the observed lowest energy resolution increases with each increase in total pressure. For example the lowest energy resolution achieved for pressures 95, 190, 380 and 760 Torr was found to be 0.175 $\pm$ 0.001 (\(p_{\text{CF}_{\text{4}}}\):10 Torr), 0.183 $\pm$ 0.002 (\(p_{\text{CF}_{\text{4}}}\):5 Torr), 0.282 $\pm$ 0.003 (\(p_{\text{CF}_{\text{4}}}\):10 Torr), and 0.424 $\pm$ 0.002 (\(p_{\text{CF}_{\text{4}}}\):40 Torr) respectively. This suggests that the energy resolution worsens as more He is added to the vessel and the total pressure increases. A similar trend of degrading energy resolution with increasing pressure was also observed in pure CF$_4$ in \autoref{sec:PureCF4}. However, for a given total pressure in \autoref{fig:He_ER}, the predictive power of the variation in energy resolution with \( \Delta V_{\text{ThGEM}}\) as the proportion of CF$_4$ changes is very low; highlighting the importance of doing these measurements experimentally. 

The results presented in \autoref{fig:He_gas_gains} and \ref{fig:He_ER} demonstrate that CF$_4$:He mixtures can produce significant charge amplification with a single ThGEM at sub-atmospheric pressures down to 95 Torr. It is believed that this success is partially due to the thickness of the ThGEM used in these measurements. Previous work with a double GEM configuration, each with a thickness of 50 $\mu$m, utilising a premixed gas bottle of CF$_4$:He with molar ratio 20:80 was not able to produce a measurable gas gain at pressures lower than 400 Torr \cite{Vahsen2014}. The smaller amplification gap likely resulted in earlier onset sparking at lower pressures. Additionally, the ability to vary the partial pressure of both constituent gases means that the mixtures tested in this paper at atmospheric pressure are more suitable for preserving the length of $^{12}$C and $^{19}$F recoils than the previously tested 20:80 atmospheric mixture. This is because the partial pressure of CF$_4$ in a 20:80 molar ratio mixture at atmospheric pressure is greater than 100 Torr.



These results also present the opportunity for optimisation of a gas mixture at atmospheric pressure. This would be beneficial as operation at atmospheric pressure can reduce the cost of a containment vessel; this consideration will be significant when scaling the detector volume up to reach meaningful cross sections. As mentioned, the lowest energy resolution achieved at atmospheric pressure can be seen to occur at 40 Torr in \autoref{fig:He_ER}. This curve also exhibits a local minima at a relatively high gas gain. It would therefore be recommended to use an atmospheric mixture with 40 Torr of CF$_4$.

Following the CF$_4$:He mixtures, pure helium was also tested at pressures of 5, 10, 20, 30, 40, 50, 95, 195, 380, and 760 Torr. However no peaks were observed in the energy spectrum before sparking occurred. This may not be surprising as He is typically mixed with a small amount of a quenching gas in gaseous ionisation detectors. Despite this result, successful charge amplification in pure He has been demonstrated before with various single/double ThGEM and WELL-ThGEM configurations; as low as 100 Torr with UV-light and 350 Torr with 5.5 MeV alpha particles \cite{Cortesi2015}. This could suggest that future measurements in pure He at low pressure would benefit from multiple amplification stages.

\section{Conclusions}
\label{sec:conclusions}

In conclusion, gas gain and energy resolution measurements were conducted in pure CF$_4$ at low pressures of 5, 10, 20, 30, 40, 50, and 100 Torr. For pressures 20 - 50 Torr, full gain curves exhibiting exponential behaviour could be achieved. Sparking occurred in both 5 and 10 Torr before a measurable gas gain could be reached and only one measurement could be made at 100 Torr. The energy resolution in pure CF$_4$ was below 0.22 in 20 - 40 Torr but appeared to increase significantly at 50 and 100 Torr. These findings suggests that lowering the pressure below 100 Torr would not only benefit directionality but also offers improvement in energy resolution. 

Following these measurements, He was added to the vessel to bring the total pressure up to 95, 190, 380, and 760 Torr. The addition of He appeared to raise the breakdown voltage and maximum stable gas gain. Above 190 Torr the maximum stable gas gain began to decrease but was still larger at 760 Torr compared to its low pressure pure CF$_4$ counterpart. The energy resolution of these mixtures worsened with increasing total pressure but some local minima were observed to coincide with relatively high gas gains, which could be used for optimisation. Besides improving the sensitivity to low WIMP masses, these results suggest that the addition of He to low pressure CF$_4$ also improves the gas gain which is beneficial to the detection of low energy recoils.

These results have demonstrated that mixtures, which prioritise a partial pressure of CF$_4$ < 100 Torr, can be successfully operated at sub-atmospheric pressures down to 95 Torr with a single ThGEM. This is lower than previously achieved with premixed ratios of CF$_4$:He and is believed to be due to the larger amplification gap used in these measurements. Additionally, the mixtures tested here are more suitable for preserving the length of both $^{12}$C and $^{19}$F recoils when compared to the previously tested 20:80 atmospheric mixture. This is due to the fact that an atmospheric mixture of 20:80 results in a partial pressure of CF$_4$ greater than 100 Torr.


These results also present the possibility of gas mixture optimisation at atmospheric pressure which is beneficial for reducing the cost of a containment vessel. This will be a significant consideration when scaling up the detector volume to reach meaningful cross sections. If an atmospheric mixture of CF$_4$:He is to be used with a ThGEM of comparable dimensions, then it is recommended that CF$_4$ contributes a partial pressure of 40 Torr to the mixture; this was found to produce the lowest energy resolution at atmospheric pressure in this case. Alternatively, when using a ThGEM of comparable dimensions under sub-atmospheric conditions, it is recommended to use a partial pressure of CF$_4$ between 20 and 40 Torr due to its stability of operation and coinciding high gain with low energy resolution.

Finally, pure He was tested with the ThGEM setup at pressures of 5, 10, 20, 30, 40, 50, 95, 195, 380, and 760 Torr but a measurable gas gain could not be achieved before sparking occurred in each case. Further work is required to test charge amplification in pure He with multistage amplification options and to understand the possible penning effects in the CF$_4$:He mixtures.

\acknowledgments
The authors would like to acknowledge support for this work from AWE and a University PhD scholarship awarded to A.G. McLean.

\end{document}